\documentclass[rmp,preprint]{revtex4}
\usepackage{graphicx}

\begin{document}
\title{X-waves Generated at Second Harmonic}
\author{Claudio Conti}
\email{c.conti@ele.uniroma3.it}
\homepage{http://optow.ele.uniroma3.it}

\author{Stefano Trillo}
\altaffiliation{Permanent address: Dept. of Engineering,
University of Ferrara, Via Saragat 1, 44100, Ferrara, Italy}
\affiliation{National Institute for the Physics of Matter, INFM- Roma Tre, Via della Vasca Navale 84, 00146 Rome, Italy}
\date{\today}

\begin{abstract}
The process of optical frequency doubling can lead,
in the undepleted regime,  to the generation of a X-wave envelope
with group velocity locked to the pump beam.
Its parameters and its angular spectrum, are directly related to
the zero- and first-order dispersive features of the nonlinear process.
This constitutes a  novel mechanism for spatio-temporal localization of light.
\end{abstract}


\maketitle
%
The fact that the {\em linear} Maxwell equations admit localized solutions dates back to the book of Stratton, \cite{Stratton41}
as pointed out more recently. \cite{Recami98}
However, this research  topics has flourished  only after Durnin et al. experiment in 1987, who reported the practical realization of non-diffracting Bessel $J_0$ beams. \cite{Durnin87}
Since then, linear waves localized in both space and time
(i.e., non-diffracting and non-dispersing) have been reported in experiments, \cite{Sonajalg1997,Saari97,Mugnai2000}, while main theoretical advances encompass the generalization of propagation invariant solutions \cite{Piestun98,Vega2000}
and the study of their finite-aperture
realizations \cite{Ziolkowski93,Shaarawi97,Stepanishen1997,Stepanishen1998},
or propagation in dispersive media \cite{Sonajalg96,Porras2001} and
waveguides \cite{Zamboni2001b}.
The aspect of superluminality \cite{Recami98,Mugnai2000}
as well as potential applications
such as femtosecond microscopy and
communication systems have been also considered. \cite{Lu99,Saari2001}

Recently, X-waves have been shown to be important
also for nonlinear optics, since they exist as propagation invariant
solutions to standard paraxial dispersive propagation models \cite{Conti2002},
whose formation is triggered by conical emission. \cite{Trillo2002}
In second harmonic (SH) generation, these wavepackets are constituted by X-shaped envelopes at SH and fundamental frequency (FF) traveling locked together, which allow to explain recent experimental results \cite{NLGW2002}.
Since these X-waves continue to exist in the linear limit
(in this sense they differ from any kind of soliton)
where the FF and SH are uncoupled,
it is natural to study the mechanism of their formation.
In this Letter we show that a X-wave at SH can be generated spontaneously via doubling of a broad FF pump beam in the undepleted stage of its propagation.

To this end, let us start by reminding that the most simple axisymmetric localized wave in a dispersive medium with refractive index $n(\omega)$
and wavenumber $k(\omega)=\omega n(\omega)/c$ ($c$ being
vacuum light velocity) can  be expressed as a superposition of Bessel beams
\cite{Sonajalg1997}
\begin{equation} \label{eq:Xformula}
E(r,z,t)=\int_0^\infty S(\omega) J_0(\kappa r) \exp[i (k_z z - \omega t)] d\omega \text{,}
\end{equation}
where $\kappa=k(\omega)\sin[\theta(\omega)]$
and $k_z=k(\omega)\cos[\theta(\omega)]$ are
the transverse and longitudinal wavenumbers, respectively.\cite{Sonajalg96}
Equation(\ref{eq:Xformula})  describes a non-diffracting
and non-dispersive beam provided that the following
{\em X-wave dispersion relation} is satisfied
 \begin{equation} \label{eq:Xdispersion}
  k_z(\omega)=\alpha + \beta_X\omega \text{,}
 \end{equation}
Indeed, Eq.(\ref{eq:Xdispersion}) entails
vanishing dispersion $\partial_{\omega}^{n} k_z=0$
at all orders $n \ge 2$, and in turn an intensity $|E|^{2}$
which depends only upon the retarded time $t-\beta_X z$,
$\beta_X=k_z'$ being the inverse wavepacket velocity.
Given the spectrum $S(\omega)$, $\alpha$
and $ \beta_X $ are the two parameters that characterize the localized wave.
In the case of vacuum, when $\alpha=0$ the spatio-temporal spectrum has an X-shape ($\kappa=\pm \omega (c^{-2}-\beta_X ^2)^{1/2}$) and,
due to invariant propagation, the far-field coincides with the near-field.
Thus  the wave retains its X-shape in the physical space $(x,y,z,t)$
and turns out to be superluminal ($c| \beta_X |<1)$
(in the  Fourier-transformable stationary regime
no general theorem about causality and Maxwell equations is violated \cite{LeviCivita}, as discussed extensively in the literature
\cite{Recami2000,Oliveira2001,Lunardi2001,Rodrigues2001}).

When $\alpha \neq 0$ the X in the spectrum is curved and subluminal propagation is allowed. \cite{Salo2001}
In general these localized waves have infinite energy, and only approximations can be realized in practice (as for non-diffracting Bessel beams \cite{Durnin87}). Solutions with finite energy, though no longer propagation-invariant, have been also reported.\cite{Oliveira2001b,Zamboni2001}
In the case of dispersive media the situation is more complicated
(the general angular spectra is given by
$\kappa^2=k(\omega)^2-( \beta_X  \omega+\alpha)^2$,
encompassing evanescent waves). However, strongly dispersive elements
such as axicon and diffraction gratings  \cite{Reivelt2000}
or computer holograms \cite{Sonajalg1997} permit to generate X-waves.

Henceforth, we consider SH generation in the undepleted pump approximation.
The paraxial and narrowband complex envelope amplitude $a=a(r,t,z)$
at SH carrier frequency $\omega_{2}=2\omega_{1}$ obeys the following equation
\begin{equation} \label{eq:MAIN}
  i\frac{\partial a}{\partial z}+i   \beta_2   \frac{\partial a}{\partial t}+
\frac{1}{2k_2} \nabla_{\perp}^2 a +
\chi A^2 e^{-i\Delta k z} = -i \epsilon a \text{,}
\end{equation}
where  $\Delta k=k_{2}-2 k_{1}$ ($k_j=k(\omega_j)$, $j=1,2$) is the nonlinear
phase mismatch, $\chi$ the usual nonlinear coefficient such that
$|a|^2$ and  $|A|^2$ are measured in $W/m^2$, and $\epsilon>0$
stands for a loss term which is included
for technical convenience, though we will take the limit $\epsilon \rightarrow 0^+$.
We also neglect second-order dispersions $k_{j}''=k''(\omega_{j})$,
since in practical cases their effects are usually small with respect to the dominant first-order dispersion measured by the group delay $\beta_1 -  \beta_2$, $\beta_j  =k'(\omega_j)$ being inverse group-velocities.
The validity of Eq.(\ref{eq:MAIN}) requires also a broad
(radially symmetric) pump beam so that its diffraction can be neglected, resulting into a pump amplitude $A=A(t-\beta_{1} z)=$ const., or in other words into a pump beam traveling undistorted at velocity $\beta_{1}^{-1}$.

\begin{figure}[h]\centerline{\scalebox{1}{\includegraphics{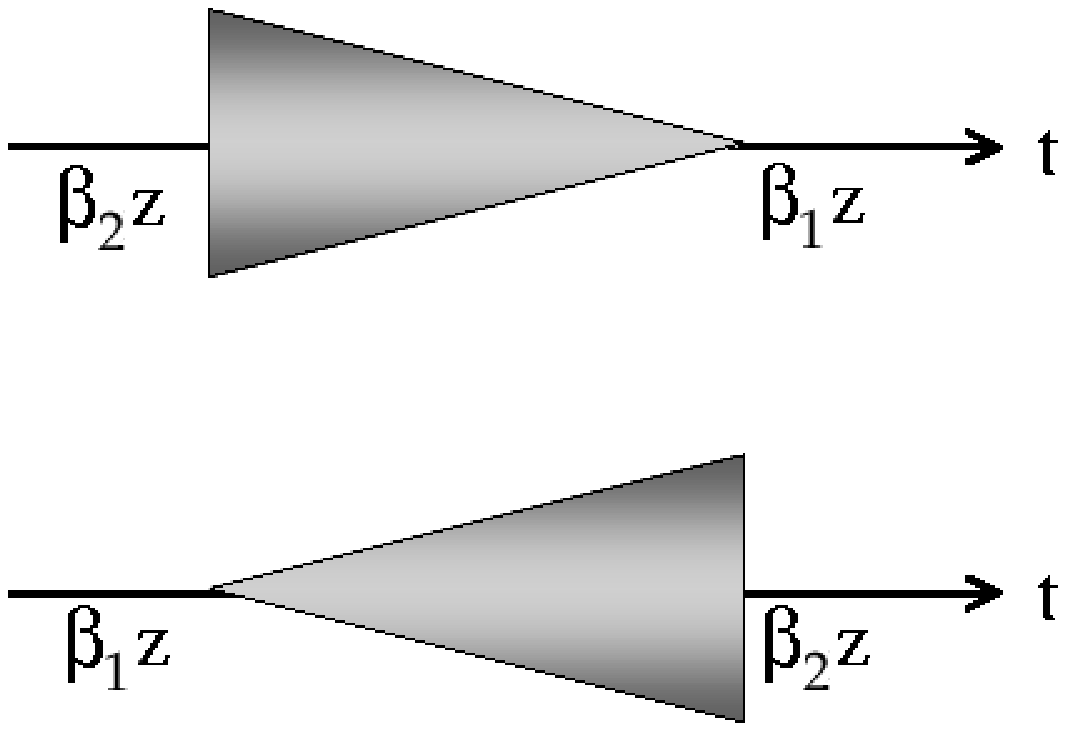}}}
\caption{Orientation (and support) of generated SH X-wave for
$ \beta_1  >  \beta_2 $ (top) and $\beta_1  <  \beta_2$ (bottom).}
 \label{figure1}
\end{figure}

Under these hypothesis, Eq.(\ref{eq:MAIN})
can be easily integrated in the spectral domain
by making use of the Fourier-Bessel representation:
$a(r,t,z)=\frac{1}{2\pi}\int_{-\infty}^{\infty}\int_0^\infty
\hat{a}(\kappa,\Omega,z) J_0(\kappa r) e^{-i\Omega t} \kappa d\kappa d\Omega$. The solution of Eq.(\ref{eq:MAIN}) in terms
of spectral variables $(\kappa,\Omega)$ reads
 \begin{equation}\label{eq:solFB}
  \hat{a}=\chi \psi \frac{\exp[i( \beta_1 \Omega   -\Delta k)z]-
\exp[i(  \beta_2  \Omega-\kappa^2/2k_2+i\epsilon)z]}
{(  \beta_1  -  \beta_2  )\Omega-\Delta k+\kappa^2/2k_2-i\epsilon} \text{.}
\end{equation}
where $\psi(\Omega,\kappa) \exp(i \Omega  \beta_1  z)$ is the
Fourier-Bessel transform of the source term $A^{2}(r,t-\beta_{1}z)$
in Eq.(\ref{eq:MAIN}), and we have considered, for sake of simplicity, the unseeded case, i.e. $a(z=0)=\hat{a}(z=0)=0$.
Note that the denominator in Eq.(\ref{eq:solFB}) represents
(for $\epsilon \rightarrow 0$) the expansion of the mismatch
$k_{z}(2(\omega_{1}+\Omega))-2k_{z}(\omega_{1}+\Omega)$.
Consistently with our hypothesis, we limit the analysis to the case $\beta_1  \neq  \beta_2  $ (when $\beta_1 = \beta_2$, the effect of $k_{j}''$ becomes relevant and will be addressed in the future).
The inverse  transform of Eq.(\ref{eq:solFB}) in the lossless limit
($\epsilon \rightarrow 0$)
can be obtained by applying residue theorems.
We find $a=0$ whenever
$t<min(  \beta_1  z,  \beta_2  z)$, or $t>max(  \beta_1  z,  \beta_2  z)$, and
\begin{eqnarray} \label{eq:mainresult}
a(r,t,z)=\frac{1}{|  \beta_1  -  \beta_2  |} \exp
\left[-i\Delta k \frac{t-  \beta_2   z}{  \beta_1  -  \beta_2 }\right] \times \nonumber \\
\int_0^\infty X(\kappa) J_0(\kappa r)
e^{i\frac{\kappa^2}{2\,k_2}\frac{t-  \beta_1  z}{  \beta_1  -  \beta_2  }}
 \kappa d\kappa
\end{eqnarray}
elsewhere (see Fig.\ref{figure1} for a pictorial representation of the support of $a$),
with $X=\psi[(-\kappa^2/2k_2+\Delta k)/(\beta_1-\beta_2),\kappa]$ determined by the pump spectrum.
Hence in the region covered by the SH (travelling at speed $1/ \beta_2$) which is emitted in the center of the FF beam,
$a$ is given by a {\em non-diffracting beam which propagates at the group
velocity of the FF}. Indeed, Eq.(\ref{eq:mainresult}) shows
that the intensity $|a|^2$ depends on $z,t$ only through the
combination $t- \beta_1  z$. Only one-half of the localized wave
is generated due to causality: the peak moves at the FF group velocity, while its costituents (the SH beam) moves at their characteristic velocity $1/  \beta_2$.
As a result, the temporal extension of the wave is equal to
$|\beta_1  -  \beta_2  |z$ (see Fig.\ref{figure1}).
This is similar to what has been theoretically predicted
for practical finite-aperture realizations of vacuum localized waves. \cite{Ziolkowski93}
As in that case, it can been shown that the $a$ field is the sum of two X-waves which
cancel each other in the non-causal regions, a form of the extintion theorem. \cite{Born80}
In the source-sink picture developed by Saari \cite{Saari2001}, only the source cone is
present (as physically reasonable).

Equation(\ref{eq:mainresult}) strongly resembles the vacuum nondiffracting beams
developed in  \cite{Salo2001}. However, since we are dealing
with a dispersive medium, we need to prove that the wave in Eq.(\ref{eq:mainresult})  belongs to the class given by (\ref{eq:Xformula}).
Consistently with the validity of Eq.(\ref{eq:MAIN}),
we make the paraxial approximation $k_z=[k(\omega)^2-\kappa^2]^{1/2} \cong k(\omega)-\kappa^2/2k(\omega)$,
and then expand $k(\omega)$ around $\omega_2$
at first-order, to end up (neglecting space-time coupling terms such as $\Omega \kappa^2$) with $k_z=k_2+  \beta_2  \Omega-\kappa^2/2 k_2$, where $\Omega=\omega-\omega_2$.
By equating this expression of  $k_z$ to that in Eq.(\ref{eq:Xdispersion}), we can express the frequency as
$\omega=\omega_{2}+(k_2- \beta_X \omega_2-\alpha-\kappa^{2}/2k_{2})/( \beta_X -  \beta_2  )$, which, after
substitution in Eq.(\ref{eq:Xformula}), yields
\begin{eqnarray} \label{eq:Xasymptotic}
E=e_2\exp \left[-i(k_2- \beta_X  \omega_2- \alpha)\frac{t-  \beta_2 z}{ \beta_X -  \beta_2  }\right] \times \nonumber \\
\int_0^\infty X'(\kappa)
J_0(\kappa r)e^{i\frac{\kappa^2}{2k_2}\frac{t- \beta_X  z}{ \beta_X - \beta_2  }}
\kappa d\kappa
\end{eqnarray}
where $X'(\kappa)$ is a suitable redefinition of the spectrum
$S(\omega)$ and $e_2=\exp[i(k_2z-\omega_2 t)]$ is the optical carrier.  The field in Eq.(\ref{eq:Xasymptotic}) has exactly
the same form of that reported in Eq.(\ref{eq:mainresult}),
once we account for the fact that the latter is an envelope
which multiply the carrier $e_2$. By comparing
Eq.(\ref{eq:Xasymptotic}) and Eq.(\ref{eq:mainresult})
we find that the parameters of the X-wave are directly related to those of the mixing process as
\begin{equation}
\beta_X =  \beta_1;\alpha=k_2- \beta_X \omega_2-\Delta k=2k_1- \beta_1  \omega_2.
\end{equation}

Therefore, our calculation shows that, for undepleted frequency doubling, the X-wave dispersion constraint (\ref{eq:Xdispersion}) is automatically satisfied and takes the form $k_z=  \beta_1 \Omega+2k_1$.
Correspondingly the angular dispersion is given by the equation
($\kappa \cong k_2 \theta$, $\theta$ being the cone angle)
\begin{equation}\label{eq:Xspectrum}
\Omega=\frac{\Delta k}{  \beta_1  -  \beta_2  }-\frac{k_2}{2( \beta_1  -  \beta_2  )}\theta^2\text{,}
\end{equation}
which is an horizontal parabola in the angle-detuning plane,
with position and convessity determined by the dispersive features of the process (group-delay and $\Delta k$). Its experimental observation would be a clear signature of the presence of angular dispersion and of the formation of SH-Xwaves.
From Eq.(\ref{eq:mainresult}) we can determine the energy of the SH 
$ \mathcal{E}(z)=2\pi{\int}_{-\infty}^{\infty}{\int}_{0}^{\infty} |a|^2 r dr dt=
\frac{\chi^2 \mathcal{F}_{FF}(\Delta k,  \beta_1  -  \beta_2  )}{| \beta_1  -  \beta_2  |}z$,
where $\mathcal{F}_{FF}={\int}_{0}^{\infty} |X|^2 \kappa d\kappa$.
$\mathcal{E}$ grows linearly during propagation, its
finiteness being a consequence of the X-wave compact support,
and hence of causality.
To make a practical example we assume the FF amplitude given
by a gaussian pulsed beam
with waist $w_F$ and duration $t_F$: $A=A_0 \exp[-(t-  \beta_1 z)^2/2t_F^2-r^2/2w_F^2]$.
In this case $\mathcal{F}_{FF}$ can be evaluated explicitly yielding
\begin{equation}
  \label{eq:energygaussian}
  \mathcal{E}=\frac{\chi^2 k_2 \mathcal{E}^2_F\,z}{2\sqrt{2\pi}t_F}
e^{-\frac{k_2 w_F^2 |  \beta_1  -  \beta_2  |}{2\sqrt{2}t_F}}e^{-\xi}
[1+erf(\sigma \xi)]\text{,}
\end{equation}
with $\mathcal{E}_F$ the FF energy,
$\xi=k_2 w_F^2 (\Delta k-\Delta \beta_X )$,
 $\Delta \beta_X =k_2 w_F^2 (  \beta_1  -  \beta_2  )^2/2t_F^2$, and
$\sigma=t_F/\sqrt{2}|  \beta_1  -  \beta_2  |k_2 w_F^2$.
Eq.(\ref{eq:energygaussian}) shows that for large $|\Delta k|$ the
energy conversion is negligibile, while the maximum efficiency is obtained
for a particular value of $\Delta k$ (in general $\Delta k \neq 0$) which can evaluated numerically.
Choosing realistic parameters \cite{NLGW2002} ($w_F=10^{-5} m$,
$t_F=10^{-13} s$, $|  \beta_1  -  \beta_2  |=45$ ps/m,
$k_2 \cong 8\pi \times 10^6 {\text m}^{-1}$), we have $\sigma \cong1$, and the maximum conversion efficiency is
attained for $\Delta k \cong \Delta \beta_X \cong |  \beta_1  - \beta_2  |/t_F$, i.e. the inverse spatial separation between FF and SH associated with  $t_F$.

At last we wish to point out that different (often equivalent) terms
are adopted to classify localized waves.
Strictly speaking, Eq.(\ref{eq:Xformula})
defines the so-called {\it Focus Waves Mode} (FWM),
a more general class than X-waves,
according to the literature.\cite{Reivelt2000}
However we are dealing with narrowband optical wavepackets,
and hence FWM are known as X-Bessel \cite{Sonajalg96,Saari97}.
On the other hand Eq. (\ref{eq:mainresult}) resembles
the {\it Focused X-waves} \cite{Saari2001}.  Furthermore the causality implies that only one-half of the double cone is generated, and thus they can be identified with the
so-called Y-, or Hankel-, waves. \cite{Salo2000}
For the sake of simplicity we used the term ``X-wave''
from the beginning.

In conclusion we have theoretically predicted that the process
of optical frequency doubling can lead to an X-wave,
traveling at the FF velocity, whose $\alpha$
parameter is directly linked to the wavevector mismatch.
For a given input FF wave, the X-wave can be
modified by acting on the mixing control knobs (e.g. crystal temperature or orientation), while the resulting
angular spectra (the far field) is given by horizontal parabolas [Eq.(\ref{eq:Xspectrum})].
This mechanism appears to be a valid alternative to the generation of optical localized-waves by means of {\em linear} techniques.

This result is also susceptible of a different interpretation.
The fact that the morphogenesy into X-waves occurs spontaneously reveals the fundamental role that they
can have in any ultrafast  parametric process (e.g. four-wave mixing or optical parametric amplification) where
the generalized phase-matching condition
produces the correlation between spatial and temporal degrees of freedom which is behind the X-shape of the wavepacket.
By studying such phenomena, the generation of new parametric sources could be conceived.

\begin{acknowledgments} We thanks P. Di Trapani and E. Recami for fruitful discussions. Financial support  from Ministry of Instruction, University and Research
(PRIN project) is acknowledged. C. C. thanks the Fondazione Tronchetti Provera for a generous grant. 
\end{acknowledgments}

\newpage

\section*{List of Figure Captions}

Fig. 1. Orientation (and support) of generated SH X-wave for
$ \beta_1  >  \beta_2 $ (top) and $\beta_1  <  \beta_2$ (bottom).



\newpage


\end{document}